\documentstyle[12pt]{article}               
\let\3=\ss \catcode`\"=\active \let"=\"     \long\def\del#1\enddel{ }
  \oddsidemargin=\evensidemargin
 \let\msk=\medskip \let\bsk=\bigskip
  
 \let\qqd=\qquad  \def\ve{\vfil\eject}

    \let\e=\varepsilon
  \let\th=\theta  
\let\l=\lambda    \let\p=\pi 
   \let\c=\chi
 
\let\Ph=\Phi  \let\Ps=\Psi   
   
\def\0{\over }    \def\1{\vec }   \def\2{{1\over2}} \def\3{{\ss}}
\def\4{{1\over4}} \def\5{\overline }   \def\6{\partial } \def\7#1{{#1}\llap{/}}
\def\8#1{{\textstyle{#1}}}        \def\9#1{{\bf {#1}}}
\def\_#1{$\underline{\hbox{#1}}$} \def\^#1{$\overline{\hbox{#1}}$}

\def\<{\langle } \def\>{\rangle }  
 
\def \({\left( } \def \){\right) }

\let\ap=\approx \let\eq=\equiv \let\ex=\times   \let\aus=\in
      \let\and=\wedge
     
\def\|#1{{}_{\bigg|_{#1}}}

\ifx\Box  \else \fi

\def\pmbf#1{\setbox0=\hbox{${#1}$}   \kern-.025em\copy0\kern-\wd0
      \kern.05em\copy0\kern-\wd0     \kern-.025em\raise.0433em\box0 }
   
  \def\cg{{\cal G}} \def\ch{{\cal H}}
   
  \def\cs{{\cal S}} 
\def\inbar{\vrule height1.5ex width.4pt depth0pt} 
\def\ifundefined#1{\expandafter\ifx\csname#1\endcsname\relax}
\makeatletter \ifundefined{new@mathgroup} {} \else 
\new@mathgroup\sffam
\new@internalmathalphabet\mathsf\sffam{cmss}{m}{n}
\def\psf{\fontfamily\sfdefault \fontseries\default@series
    \fontshape\default@shape\selectfont\mathsf}
\fi
\def\ZZ{\relax{\sf Z\kern-.4em \sf Z}}  \def\IR{\relax{\rm I\kern-.18em R}}
\def\IN{\relax{\rm I\kern-.18em N}} \def\IP{\relax{\rm I\kern-.18em P}}
\def\IQ{\relax\,\hbox{$\inbar\kern-.3em{\rm Q}$}}
\def\IC{\hbox{\,$\inbar\kern-.3em{\rm C}$}}
\def\citen#1{\if@filesw \immediate\write \@auxout {\string\citation{#1}}\fi%
\@tempcntb\m@ne \let\@h@ld\relax \def\@citea{}%
\@for \@citeb:=#1\do {\@ifundefined {b@\@citeb}%
    {\@h@ld\@citea\@tempcntb\m@ne{\bf ?}%
    \@warning {Citation `\@citeb ' on page \thepage \space undefined}}%
    {\@tempcnta\@tempcntb \advance\@tempcnta\@ne
    \setbox\z@\hbox\bgroup\ifcat0\csname b@\@citeb \endcsname \relax
       \egroup \@tempcntb\number\csname b@\@citeb \endcsname \relax
       \else \egroup \@tempcntb\m@ne \fi \ifnum\@tempcnta=\@tempcntb
       \ifx\@h@ld\relax \edef \@h@ld{\@citea\csname b@\@citeb\endcsname}%
       \else \edef\@h@ld{\hbox{--}\penalty\@highpenalty
	      \csname b@\@citeb\endcsname}\fi
    \else \@h@ld\@citea\csname b@\@citeb \endcsname \let\@h@ld\relax \fi}%
 \def\@citea{,\penalty\@highpenalty\hskip.13em plus.13em minus.13em}}\@h@ld}
\def\@citex[#1]#2{\@cite{\citen{#2}}{#1}}%
\def\@cite#1#2{\leavevmode\unskip
  \ifnum\lastpenalty=\z@\penalty\@highpenalty\fi
  \ [{\multiply\@highpenalty 3 #1
  \if@tempswa,\penalty\@highpenalty\ #2\fi}]}   
\makeatother 

\def\beq{\begin{equation}} \def\eeq{\end{equation}} \def\eql#1{\label{#1}\eeq}
\def\bea{\begin{eqnarray}} \def\eea{\end{eqnarray}} 
\def\fnote#1#2{\begingroup\def\thefootnote{#1}\footnote{#2}
	   \addtocounter{footnote}{-1}\endgroup}

\def\plb#1 #2 {Phys. Lett. {\bf B#1} #2 }
\def\phr#1 #2 {Phys. Rep. {\bf  #1} #2 } 
\def\npb#1 #2 {Nucl. Phys. {\bf B#1} #2 }
\def\aph#1 #2 {Ann. Phys. {\bf #1} #2 }  \let\ap=\aph
\def\jmp#1 #2 {J. Math. Phys. {\bf #1} #2 }
\def\prd#1 #2 {Phys. Rev. {\bf D#1} #2 }
\def\prl#1 #2 {Phys. Rev. Lett. {\bf #1} #2 }
\def\rmp#1 #2 {Rev. Mod. Phys.  {\bf #1} #2 }
\def\zpc#1 #2 {Z. Phys. {\bf #1C} #2 }
\def\cmp#1 #2 {Comm. Math. Phys. {\bf #1} #2 }
\def\mpl#1 #2 {Mod. Phys. Lett. {\bf A#1} #2 }
\def\ijmp#1 #2 {Int. J. Mod. Phys. {\bf A#1} #2 }

\def\VS#1#2{\vrule height #1mm depth #2mm width 0pt}
\def\TS{@{~~\VS{5}0}}  \def\TL{\VS02} 

\def\naive{na\"\i ve} \def\[{\left[} \def\]{\right]} 
\def\ng{n_{27}} \def\na{n_{\overline{27}}} 
\def\tbf#1:{{\noindent\bf #1:}}            \long\def\new#1\endnew{{\bf #1}}
   
  \def\PD{Poincar\'e duality}
\def\LG{Landau--Ginzburg}   \def\LGO{\LG\ orbifold}   \def\CY{Calabi--Yau}
\def\CFT{conformal field theory} \def\CFTs{conformal field theories}

\let\Ph=X  \let\Ps=Y

\def\figuresonly{\pagestyle{empty}\figa\ve\figb\ve\figc\end{document}}
\long\def\old#1\endold{{\small #1}}         \def\oldansw{o } \def\cutansw{c }
\newcount{\npre} \npre=1  \def\negansw{s } \def\figansw{f } \def\textansw{t }
\def\ifpre{\ifnum\npre=1 } \def\ifsub{\ifnum\npre=0 }        \def\cut#1{#1}
\def\askversion{\message{
Preprint (p) / submit (s) / text only (t) / figures only (f):  (p/s/t/f)? }
    \read-1 to\answ \ifx\answ\negansw \npre=0 \else \npre=1 \fi
    \ifx\answ\figansw { } \else \def\figuresonly{ }   \fi
    \ifx\answ\oldansw \def\old##1\endold{{\small ##1}}\fi
    \ifx\answ\textansw \npre=2 \else \message{
Cut figures (use 'c' in case of memory problem):  (c/n)? }
    \read-1 to\answ\ifx\answ\cutansw \def\cut##1{}\npre=7\fi\fi \figuresonly }

\def\bpic{\begin{picture}} \def\epic{\end{picture}} \thicklines
\def\lab#1)#2#3{\put#1){\makebox(0,0)[#2]{\small #3}}}
\def\putlin#1,#2,#3,#4,#5){\put#1,#2){\line(#3,#4){#5}}}
\def\putvec#1,#2,#3,#4,#5){\put#1,#2){\vector(#3,#4){#5}}}

\newcount{\vmul} \newcount{\hdiv} 
\newcount{\wi}   \newcount{\he}   
\newcount{\hq}   \newcount{\vq}   
\newcount{\hoff} \newcount{\auxc} \newcounter{figco}   \def\npt{\circle*{2}}

\def\vlline{\put(-3,0){\line(1,0)6}}      
\def\vlright#1{\put(6,0){\makebox(0,0)[l]{\scriptsize #1}}}
\def\putvl#1{\mbox{\bpic(0,0)\funit=1pt\vlline\vlright{#1}\epic}}
\def\putvm#1{\mbox{\bpic(0,0)\funit=1pt\vlline\epic}}   
\def\vlab#1{\vq=#1\multiply\vq by\vmul \put(-\hoff,\vq){\putvl{#1}}
	    \put(\hoff,\vq){\putvm{#1}} }
\def\vmark#1{\vq=#1\multiply\vq by\vmul \put(-\hoff,\vq){\putvm{#1}}
	    \put(\hoff,\vq){\putvm{#1}}  }

\def\hlline{\put(0,-3){\line(0,1)6}}     
\def\hltop#1{\put(0,6){\makebox(0,0)[b]{\scriptsize #1}}}
\def\puthl#1{\mbox{\bpic(0,0)\funit=1pt\hlline\hltop{#1}\epic}}
\def\hlab#1{\hq=#1\divide\hq by\hdiv \put(\hq,0){\puthl{#1}} }    
\def\hlabo{\put(0,0){\mbox{\bpic(0,0)\funit=1pt\put(0,-3){\line(0,1)3}\epic}}}

\def\Vpt#1,#2){\hq=#2\advance\hq by -#1 \multiply\hq by 2 \divide\hq by\hdiv
	       \vq=#1\advance\vq by #2 \multiply\vq by\vmul\put(\hq,\vq){\npt}}
\def\Vplo#1{\vbox{\hdiv=2\vmul=1 \figsca \auxc=\he \multiply\auxc by\vmul
    \hoff=\wi\divide\hoff by2 \stepcounter{figco}\message{[Fig. \arabic{figco}}
    \begin{center}\let\.=\Vpt \bpic(\wi,\auxc)(-\hoff,0) \figlab #1 \hlabo
    \put(-\hoff,0){\framebox(\wi,\auxc){}} \epic \\[5mm]
    Fig. \arabic{figco}: \figcap \end{center}} \vfil \message{]}}
\def\figsca{\unitlength=1.1pt \wi=500 \he=400} \let\funit=\unitlength

\begin{document}
\def\hannover{ITP--UH--20/93} \def\cern{CERN-TH.6931/93}
{\hfill\cern\vskip-9pt \hfill\hannover\vskip-9pt \hfill
hep-th/9307145}
\vskip 15mm \centerline{\hss\Large\bf
       ADE string vacua with discrete torsion \hss}
\begin{center} \vskip 8mm
       Maximilian KREUZER\fnote{*}{Address after October 1, 1993:
       Institut f"ur Theoretische Physik, Technischen Universit"at Wien,
       Wiedner Hauptstra\3e 8--10, A-1040 Wien, AUSTRIA}
\vskip 3mm
       CERN, Theory Division\\
       CH--1211 Geneva 23, SWITZERLAND
\vskip 6mm               and
\vskip 3mm
       Harald SKARKE\fnote{\#}{e-mail: skarke@kastor.itp.uni-hannover.de}
\vskip 3mm
       Institut f"ur Theoretische Physik, Universit"at Hannover\\
       Appelstra\3e 2, D--3000 Hannover 1, GERMANY

\vfil                        {\bf ABSTRACT}                \end{center}

We complete the classification of (2,2) string vacua that can be constructed by
diagonal twists of tensor products of minimal models with ADE invariants.
Using the \LG\ framework, we compute all spectra from inequivalent models
of this type. The completeness of our results is only possible by
systematically avoiding the huge redundancies coming from permutation
symmetries of tensor products. We recover the results
for (2,2) vacua of an extensive computation of simple current invariants by
Schellekens and Yankielowitz, and find 4 additional mirror pairs of spectra
that were missed by their stochastic method.
For the model $(1)^9$
we observe a relation between redundant spectra and
groups that are related in a particular way.

\vfil\noindent \cern\\ \hannover\\ July 1993 \msk
\thispagestyle{empty} \newpage
\setcounter{page}{1} \pagestyle{plain}
\ifsub \baselineskip=20pt \else \baselineskip=14pt \fi

\section{Introduction}

In addition to the geometrical compactification on \CY\ manifolds, it soon
became clear that abstract \CFTs\ can be used to construct perfectly sensible
4-dimensional string vacua.
When starting from the standard heterotic construction with gauge group
$SO(10)\ex E_8$,   there are basically three approaches
for getting a modular invariant string theory with
space--time supersymmetry.
For tensor products of $N=2$ minimal models, Gepner used an orbifold-type
construction~\cite{g}.
Aiming at twisted \LG\ models, Vafa~\cite{v} could prove modular invariance
and charge integrality for a very general class of orbifolds.
Alternatively, Schellekens and Yankielowitz~\cite{sy} used simple currents in
their construction of consistent string vacua.
Of course,  there is a large overlap between these methods.
In particular, Gepner's models are all contained in the other two classes.
Since the symmetries of $N=2$ minimal models are generated by simple
currents, the last approach is the most general one using only diagonal
symmetries of ADE models~\cite{ks}.
Permutation symmetries of identical factors~\cite{fkss}, on the other hand,
can be modded in a simple and general way using the~\LG\ framework\cite{iv},
but are outside the class of the simple current invariants.
For diagonal symmetries of ADE models, our results indicate that, at least for
(2,2) vacua, these two approaches are actually equivalent.

In the present work we complete the construction of string vacua with a
global (2,2) world-sheet supersymmetry that can
be obtained by diagonal twists of tensor products of $N=2$ minimal models
with ADE modular invariants,
including the most general discrete torsions.
Our motivation for focusing on this class of models is twofold: First, there
have been extensive, but incomplete, computations in the
literature~\cite{sy,fkss,lr,ls,fiqs,aas}, which
allow a comparison of results. Here our main contribution is to systematically
avoid redundancies in the crucial cases for which a \naive\ count of the
number of possibilities seems to render completeness elusive.
Furthermore, concerning this huge degeneracy due to permutation symmetries,
the ADE models obviously also contain the crucial
cases of the much larger class of \LG\ string vacua. Thus our work is also
a necessary first step for a complete enumeration of the corresponding
large set of spectra~\cite{lgt}.

As we use the \LG\ description of $N=2$ minimal models for the calculation of
spectra, we first recall some results on \LGO s  and describe the most
general structure of spectra that we have to expect.
In this framework simple formulas exist for the charge degeneracies
of the chiral primary fields (or the Ramond ground states).%
\footnote{We use the term `spectrum' for the non-singlet part of the massless
    string modes.
    The numbers of $E_6$ singlets could, if necessary, easily be obtained
    by just replacing (or extending) the subroutine for the computation of
    spectra in our computer program, which is only a small part of the code.}
It turns out that it is sufficient to consider only the well-known 168
potentials that describe pure diagonal invariants~\cite{lr}, plus the 12
potentials that in addition contain one or more $E_7$ invariants. In fact,
we can argue that we can also omit the more symmetric of the two tensor
products with 8 factors. The remaining models then only require modest
computation time, except for the unique product of 9 factors.

It is thus only in the case of the model $(1)^9$, i.e. the tensor product of
9 minimal models at level 1, that we need to eliminate the redundancy coming
from permutation symmetries. This model is treated in some detail in
section~3. We first describe a simple method that allows us to eliminate
all redundancies in the twist groups as well as in the corresponding allowed
values of the discrete torsions. We end up with about $10^5$ orbifolds
that are all inequivalent from the conformal field theory point of view.
The massless spectra, however, are still highly degenerate.
In section 4 we present and discuss our results. As expected
for our complete calculation,
we find perfect mirror symmetry for this class of (2,2) models.
Studying the explicit group structures, we observe
that a group only produces new spectra if its projection in the (untwisted)
chiral ring is irreducible in a certain sense.
If true in general, this could explain most of this remaining degeneracy and
provide an efficient computational tool.

\section{\LGO s and their spectra }

The \LG\ description of superconformal field theories relies on the
assumption that an $N=2$ supersymmetric theory with a quasi-homogeneous
superpotential $W(\l^{n_i}\Ph_i) = \l^dW(\Ph_i)$
will flow to an $N=2$ superconformal model at the renormalization group fixed
point~\cite{mvw}.
For transverse polynomials $W$, the central charge $c=3\sum_i(1-2q_i)$
can be computed from the charges $q_i=n_i/d$ of the superfields $\Ph_i$
\cite{lvw}. The potentials with $c<3$, which are
easily enumerated, are in one-to-one correspondence with the $N=2$
supersymmetric minimal models, including all non-diagonal modular invariants:
\bea &&~A_n:\;W=\Ph^{n+1},~  \qqd D_n:\;W=\Ph^{n-1}+\Ph\Ps^2,\\
     &&E_6:\;W=\Ph^3+\Ps^4, \qqd E_7:\;W=\Ph^3+\Ph\Ps^3,\qqd
     E_8:\;W=\Ph^3+\Ps^5.   \label E                       \eea
The chiral ring structure and in particular the charge degeneracies,
as derived from the \LG\ description, coincide with those from conformal field
theory. Therefore the uniqueness of the ADE classification implies that these
models must coincide if appropriate fixed points exist.
In terms of the above superfields the discrete symmetries of the minimal
models are realized by simple phase symmetries.
The minimal model $(k)$ at level $k$ with the diagonal invariant, which has
a discrete $\ZZ_{k+2}$ symmetry, thus corresponds to $A_{k+1}$.
The non-diagonal models $(k)_D$ exist at even levels $k\aus2\ZZ$ and
are described by the potentials $D_{ \2 k+2}$,
and finally there are the exceptional invariants $E_6$, $E_7$ and $E_8$ at
levels $k=10,16,28$.

By inspection of the superpotentials (\ref E), one sees that $E_6$ is a direct
product of $A_2$ and $A_3$, and that $E_8$ is a direct product of $A_2$ and
$A_4$. This has also been shown directly for the corresponding \CFTs, provided
that Ramond and Neveu--Schwarz states are paired only among themselves,
i.e. for all models with at least (1,1) superconformal invariance on the
world sheet \cite{ex}.
It is also well-known that $A$ and $D$ invariants are $\ZZ_2$
orbifolds of one another.\footnote{~From the $D$ invariant we can get back
to the $A$ invariant by modding the ``quantum'' $\ZZ_2$~\cite{q};
in case of extended chiral algebras this is less obvious in the simple current
formalism than for the \LG\ description.}
As we want to twist these models by all phase symmetries using the
full set of consistent discrete torsions, the $D$ invariants
should not yield anything new.
Indeed, we checked in a large number of cases that the set of spectra
we obtained for models that are related by exchange of $A$ and $D$ invariants
coincide exactly (this is true also in the presence of $E_7$).
Therefore in principle we only need to consider the 168 different combinations
of minimal models with central charges adding up to 9 and
with diagonal invariants,
which were listed in refs.~\cite{lr,ls}, and the additional 12 cases
where one or more of the models at level 16 come with the $E_7$ invariant.

For space--time supersymmetric vacua we require that all
(left) charges are integral. To this end we orbifoldize these models using
all diagonal symmetries and discrete torsions that are consistent with
(2,2) superconformal invariance and integral charges~\cite{iv}.
For the construction of the phase symmetries we use the procedure
that is already discussed in \cite{aas}.

Our calculation of the spectra is based on the results of refs. \cite{v,iv}:
The chiral ring of a \LGO\ can be computed by summing over all twisted sectors,
taking into account that the chiral states of each sector correspond to the
reduced \LG\ model that is obtained by setting to 0 all fields that are
twisted by a non-trivial phase.
Thus we only need the charges and the transformation properties of the
twisted vacua $|h\>$ in the $(c,c)$ ring (Ramond ground states and anti-chiral
states can be obtained from this by spectral flow).
Let the field $X_i$
transform with a phase ~$\exp 2\p i\th_i^h$~ under the
group element $h$. Then
the semi-classical analysis of refs.~\cite{v,iv}
yields the left/right charges
\beq Q_\pm = \sum_{\th_i\not\aus\ZZ}~(\8\2-q_i)\pm(\th_i^h-\8\2) \eql{Charge}
and the transformations
\beq g|h\>=(-1)^{K_gK_h}\e(g,h){\det g_{|_h}\0 \det g}|h\>         \eql{Phase}
for commuting group elements $g$ and $h$. Here
$K:\; g\to K_g\aus\{0,1\}$ is a homomorphism of the group into $\ZZ_2$, which
determines the signs of the actions of the symmetry generators on the
Ramond sector.
The discrete torsions $\e(g,h)$ have to be multiplicative in both entries
and must fulfil $\e(g,h)\e(h,g)=\e(1,g)=1$.
The projector to integral left charges $j=e^{2\p i J_0}$ always has to
be part of the twist group.
In order to get a (2,2) theory with integral left and right charges,
we must require $\e(j,g)=(-1)^{K_gK_j}\det g$ and $(-1)^{K_g}=\det g$
(thereby restricting the possible group elements to having determinants
$\pm 1$).

We denote by $p_{ij}$ the number of states with $U(1)$ charges
$(q_L,q_R)=(i,j)$, and we now want the find the most general form of these
numbers for (2,2) vacua that can come from \LGO s. First, charge conjugation
and spectral flow imply the \PD\ $p_{i,j}=p_{D-i,D-j}$, i.e. for $D=3$
we have $p_{11}=p_{22}$ and $p_{12}=p_{21}$.
These numbers determine the numbers of generations and anti-generations.
One can show that there are further restrictions for states with vanishing
left or right charge, namely $p_{0,i}=p_{0,D-i}$ and $p_{i,0}=p_{D-i,0}$
\cite{lgt}.
We also know that
$p_{33}=p_{30}=p_{03}=p_{00}=1$ (of course this is exactly
what we wanted to achieve with the restrictions on $\e$ and $K$).
The remaining freedom is in
$p_{01}=p_{02}$ and $p_{10}=p_{20}$. These numbers can only assume the values
0, 1 or 3. From \CFT\ we know that in the context of the
heterotic string a positive $p_{01}$ or $p_{10}$ implies an extension of the
gauge group to (at least) $E_7$ or extended space--time supersymmetry.
As this excludes the possibility of having chiral fermion generations,
$p_1=p_{01}+p_{10}$ can be non-zero only if the
Euler number $\c=2(p_{12}-p_{11})$ vanishes.
In fact, if $p_{01}=3$ or $p_{10}=3$ we always
find that $p_{ij}$ factorizes into left and right contributions,
i.e. $p_{ij}=p_{i0}p_{0j}$;
otherwise this product only gives a lower bound for $p_{ij}$.

In \cite{iv} the possibility of twisting by group actions with negative
determinants in case of odd $d$ was excluded because of the constraints
on the discrete torsions with $j$.
We think,
however, that this is too restrictive for the following reasons:
We can always make $d$ even by adding a 
trivial field $X_0$ with a contribution $X_0^2$ to the superpotential.
Even without resorting to trivial fields, it is easy
to see that the fields in the Ramond sector are quantized in
units of $1/2d$ rather that in units of $1/d$.
(Accordingly, the simple current that corresponds to $j$ is of order $2(k+2)$
for minimal models at odd levels $k$.)
Both arguments show that, for odd $d$,
it is sensible to assume that $j$ is of order $2d$ rather than of order $d$,
implying that the torsion constraints no longer exclude a negative determinant.
We do, however, agree with ref.~\cite{iv} in the conclusion that such twists
can be ignored, because they cannot yield anything new
(see also ref.~\cite{lgt}).

When combined, these results are very useful: We can omit
the model $(1)^7(4)$ from our computation, which would be the
worst case concerning redundancies, except for $(1)^9$.
In a first step, we can replace $(1)^7(4)_A$ by $(1)^7(4)_D$.
The phase symmetries of $X_1^3+X_1X_2^2$, however, are contained in the
symmetries of $X_1^3+X_2^3$ except for a $\ZZ_2$ factor with negative
determinant. That generator can be omitted without losing any spectra,
and thus all possible spectra are already present in the model $(1)^9$.
(The authors of ref.~\cite{ex} arrive at a similar conclusion within a
rather different framework.)

For simple types of superpotentials, in particular the present case of
ADE models, it is quite straightforward to implement
the explicit sum over the chiral ring and the
projection to invariant states in a computer program.
An alternative, which is independent of the singularity type of $W$
and therefore does not require the construction of a basis of the chiral ring,
is based on the index formula
\beq -\c={1\0|G|}\sum_{gh=hg}(-1)^{N+K_gK_h+K_{gh}}\e(g,h)
     \prod_{\th_i^g=\th_i^h=0} {n_i-d\0n_i}                           \eql X
for the Euler number \cite{iv}. A similar formula can be derived for
the dimension $\5\c\eq\sum_{ij}p_{ij}$ of the chiral ring
\beq \5\c={1\0|G|}\sum_{gh=hg}(-1)^{N_h+K_gK_h+K_{g}}\e(g,h)
     \prod_{\th_i^g=\th_i^h=0} {n_i-d\0n_i},                          \eql{Xb}
where $N_h$ denotes the number of
fields that are invariant under $h$ \cite{aas}.
Together with $p_1$ these two quantities allow computation of $p_{11}$ and
$p_{12}$.
Contributions to $p_{10}$ and $p_{01}$, which must be zero for $\c\ne 0$,
can only come from twisted ground
states~\cite{nms}, so we do not need
any explicit knowledge of the ring
structure for the calculation of the complete non-singlet spectrum.
As a check we compared the results of the two programs in many
examples. The algorithm that constructs the chiral ring
turns out to be much faster.

\section{Redundancies and the model $\bf(1)^{9}$}

If all factors in the tensor product are different, the algorithm for
computing subgroups that we gave in ref.~\cite{aas}
constructs all inequivalent subgroups of an abelian symmetry group exactly
once. For each such group we then have to generate all antisymmetric
matrices that satisfy the quantization conditions appropriate for discrete
torsions~\cite{dt}.
In case of permutation symmetries, however, there
can be a huge redundancy both in the groups and in the discrete torsions
that are generated. In the present case of the model $(1)^9$, for example,
we would already get $3^{21}$ different torsion matrices for the single
maximal symmetry group with determinant 1, which is $(\ZZ_3)^8$, if the
torsion constraint with $j$ is taken into account.
Unfortunately, finding an algorithm that directly
constructs exactly the inequivalent cases is very difficult in general.

{}From a pragmatic point of view, however,
we do not need a direct construction, because we can let the
computer sort out the redundant cases.
Indeed, there is a
conceptually very simple algorithm to avoid any redundancies,
and  with refinements it also works sufficiently fast.
The idea is to write the group action as an array and to impose an ordering
(lexicographical, for example).
Then we consider all the transformations of this array induced by permutations
of identical  factors
in the tensor product and keep a particular group action
only if it is minimal in the set of equivalent actions. By
remembering the automorphisms that are found as a by-product in this process
we can then also eliminate the redundant torsions in a similar way.

Generating all the groups and then all permutations for each
of them would of course take much too long; but we can organize the
generation of groups in such a way that we only get a modest redundancy in
the first step and at the same time only have to check for a subset of the
permutations that have a chance to give a ``smaller'' group.
Fortunately, we only had to implement this procedure for the single model
$(1)^9$, for which the whole computation took a few days on a workstation.
The elimination of redundancies for that model is illustrated in table~1:
Altogether, the \naive\ counting overestimates the number of models by more
than 5 orders of magnitude.
The only other critical case could be omitted, as we discussed in section~2.
The remaining 178 models together also needed a few days. We now describe
in more detail how the above concept can be applied to the model with the
maximal permutation symmetry $\cs_9$.

We first introduce a convenient notation for a general diagonal symmetry
group. Such a group of rank $r$ can be represented by a matrix
\beq      \cg^{(r)}=\({\bf 1}_r,M\)     \eql G
of integers modulo 3, where ${\bf 1}_r$ is the unit matix and $M$ is
an $r\ex(9-r)$ matrix. Here we already used the freedom to redefine
generators and to permute the identical factors in the tensor product,
i.e. the columns of the $r\ex9$ matrix $\cg^{(r)}$. The determinant condition
is taken into account by requiring that the sum of the numbers in each
line is a multiple of~3; furthermore,
as the generator $j$
has to be contained, the last line can be computed in terms of the remaining
ones by requiring that the sum of the numbers in each column is 1 modulo~3,
where of course $1\le r\le8$.

Now we define the normal form of a group as the smallest number whose
$r(9-r)$ digits, which are 0, 1 or 2, are the lines of a matrix $M$ in an
equivalence class
(we let this number begin with the last line). The remaining
permutations that are consistent with the choice (\ref G) induce three
types of transformations of the matrix $M$: Permutations of the last $9-r$
factors simply permute the columns of $M$. Permutations of the first $r$
factors, on the other hand, induce a permutation of the lines of $M$. It is
straightforward to make efficient use of this freedom already in the
construction of candidate matrices $M$ by choosing their entries
in such a way that no permutation of a
line can be `smaller' than the last one and that
within each line the permutations of columns that are not yet fixed by the
previous lines are taken into account.
When we finally check for possible smaller matrices,
we only need to consider those permutations of
lines that replace the last one by a line consisting of the same digits.
Furthermore, if we keep for each line the information about the residual
permutation symmetry, we can automatically use an optimal permutation
of the columns and thereby considerably reduce the number of
transformations that have to be checked.
The whole procedure of generating all inequivalent groups then only takes a
fraction of a second.

\begin{table}   \centering \caption{
Numbers of (inequivalent) groups and torsions for the model $(1)^9$.}
 \vspace{4 mm} \def\.{\8{\times}}
\begin{tabular}{||c|cccccccc|c\TS||} \hline\hline
rank         & 1 & 2 &  3 &  4 &  5 &  6 & 7 & 8 & together \TL\\ \hline
groups   & 1 & 1093 & 99463 & 925771 & 925771 & 99463 & 1093 &1&$2.1\.10^6$\\
inequiv. & 1 & 5 & 21 & 49 & 49 & 21 & 5 & 1 & 152 \\
spectra  & 1 & 5 & 14 & 22 & 22 & 14 & 5 & 1 &  23 \TL\\ \hline
torsions & 1 & 1093 &$\!298389\!$&$\!2.5\.10^7\!$&$\!6.7\.10^8\!$&
    $\!5.9\.10^9\!$&$\!1.6\.10^{10}\!$&$\!1.1\.10^{10}\!$&$3.3\.10^{10}$\\
inequiv. & 1 & 5 & 43 & 370 & 3517 & 20604 & 48285 & 30867 & 103692 \\
spectra  & 1 & 5 & 20 & 29 & 31 & 31 & 31 & 31 & 31 \TL \\
\hline\hline\end{tabular}\bsk \end{table}

Finally, there are $9!/(r!(9-r)!)$ possibilities for choosing subsets of $r$
factors to become the first $r$ factors in the tensor product.
Together with the permutations within each sector, they
generate the full symmetric group $\cs_9$.
Since we require that the matrix of generators is of the form $({\bf 1},M)$,
such a choice is only consistent if the square matrix consisting of
the corresponding
$r$ columns can be transformed into the unit matrix, i.e. is invertible.
For a given matrix $M$, this excludes, in general, some of the
choices; for each of the remaining ones
we have to perform the matrix inversion and then repeat
the procedure described in the last paragraph. Using all these
transformations to check for the normal form of $M$, we end up with 152
inequivalent groups (see  table~1). The calculation of all
orbifolds without torsion for the model $(1)^9$ now takes 9 seconds
(instead of almost one week if we ignore permutation symmetries~\cite{aas}).

The described procedure to avoid redundancies becomes crucial  when we
include discrete torsions, as the number of consistent torsions per group
and the computation time per orbifold both grow exponentially with the
rank of the group. Here we have to generate all anti-symmetric
$r\ex r$ matrices with the constraint that the numbers in each line add up
to 0 modulo 3 (with our conventions, this is necessary to have $\e(j,g)=1$
for all group elements $g$).
We again represent these matrices
as single numbers $T$, in this case with $r(r-1)/2$ digits.
Then we consider all automorphisms of the group and the transformations of
the torsion matrix they induce.
As soon as any of these transformations makes $T$ smaller we can omit that
torsion. In practice we follow a strategy analogous to the one described
above: We use  permutations  of the generators,
i.e. permutations of the lines of $M$ that can be compensated by
permutations of its columns,
to restrict the torsions that are generated in the first place. In the
crucial case of rank 8 we even have the full symmetry $\cs_8$.
Fortunately, all automorphisms of the group have already been generated
as a by-product in the check for redundant groups, and can now be used
to eliminate all redundant torsions.%
\footnote{For rank 8 there are $9!$ automorphisms, which would require a
   considerable amount of memory. Furthermore, a check for redundant
   groups is not necessary in this case; we therefore explicitly generate
   the automorphisms, or rather the obvious transformations they induce,
   in the torsion part of our program.}

\section{Results and discussion}

We calculated all (2,2) spectra coming from diagonal twists for all
potentials consisting of $A$-, $D$- and $E$-type contributions, with the
two  exceptions $(1)^7(4)_A$ and $(1)^7(4)_D$.
As the diagonal invariant and the $D$ invariant are $\ZZ_2$ orbifolds
of one another, the respective sets of spectra that can arise when
arbitrary twists and torsions are used, should be identical.
We verified for many examples, as well as for the complete set of spectra,
that exchanging $A$ and corresponding $D$-type contributions indeed
does not change the result.
This allowed us to omit the $A$-type potential corresponding to $(1)^7(4)$;
for the corresponding $D$-type invariant we can show that its spectra
are contained in  those of
the  model $(1)^9$. For the latter model we constructed
the complete set of orbifolds, which was possible by eliminating all
redundancies as described in section~3. Thus we indeed enumerated all
(2,2) spectra for ADE orbifolds.

All in all, we got only 943 different spectra (to be compared, for example,
with the 3798 spectra found in \cite{aas} for general \LGO s without torsion).
Our results are most easily represented graphically.
Figure 1 shows all spectra that we found in a plot that displays
$\ng+\na$ over the Euler number. It turns out that spectra coming
from potentials that contain the $E_7$ term $X_1^3+X_1X_2^3$ are all in
the region of a relatively low total number of non-singlets. Figure 2,
which is
an enlarged view of this region, contains all these spectra.
Spectra that come only from potentials that
do not contain the $E_7$ term $X_1^3+X_1X_2^3$ are represented by small dots,
spectra coming only from potentials containing that term by large dots,
and spectra occurring for both types of potentials by circles.
Figure 3 displays in a similar way all spectra that cannot be obtained
without the use of discrete torsion. Here large dots represent spectra
that cannot come from any Landau--Ginzburg orbifold without torsion \cite{aas},
whereas circles represent those of the remaining spectra that do not
come from ADE models without torsion. All other ADE spectra, as far as they are
in the range of the plot, are represented by small dots.

As we expected, figures 1 and 2 show complete mirror symmetry.
This symmetry under the exchange of $\ng$ and $\na$ has been shown to be
present for orbifolds of $A$-type potentials without torsion by
Greene and Plesser~\cite{gp}. More recently, their construction was
extended to more general types of potentials, including $E_7$, by
Berglund and H\"ubsch~\cite{bh}.
Using the properties of quantum symmetries, it should be straightforward
to extend these constructions to cover the case of arbitrary discrete torsions.
This is indeed confirmed by our results, and the complete mirror symmetry
for $D$-type potentials is then a consequence of the $A-D$ equivalence.%
\footnote{In ref.~\cite{bh} the mirror spectra for $D$-type potentials are
constructed in terms of the corresponding diagonal invariant; in our case of
arbitrary torsion, however, this makes no difference and we find perfect
symmetry for the spectra of each potential.}

It is a priori not obvious that we are dealing with the same class of (2,2)
models as were constructed with simple current techniques
by Schellekens and Yankielowicz~\cite{sy}.
It turns out, however, that we recover exactly all their (2,2) spectra,
and find four additional mirror pairs.
These four spectra come from only two different potentials:
In $(1)(16)^3$, with no $E_7$ invariant,
we find the mirror pairs $\{p_{11},p_{12}\}=\{47,29\}$, $\{101,11\}$ and
$\{110,2\}$ with Euler numbers $\pm36$, $\pm180$ and $\pm216$
(this is the model \#117 in the list of ref.~\cite{lr});
the spectrum with $\chi=-180$ was already found in ref.~\cite{fksv}.
In addition, for $(2)(4)(12)(82)$, i.e. the model \#127,
we find $\{p_{11},p_{12}\}=\{39,33\}$ with $\chi=\pm12$. This
indicates that the two approaches indeed cover the same space of (2,2) vacua.
The absence of the above four mirror pairs in the lists of ref.~\cite{sy}
might be due to the stochastic nature of their search.
On the other hand, a complete classification of simple current invariants
became available only recently~\cite{ks,gs}. Therefore it is conceivable that
these spectra require a complicated invariant, which was not yet included in
the ensemble that was scanned in ref.~\cite{sy}.

\begin{table}   \centering \caption{
The 31 spectra for the model $(1)^9$ and the respective minimal ranks $r$
($r_0$) of twist groups with (without) torsion. The last two columns of
spectra have $\c=0$, and the last one is left--right asymmetric.} \vspace{4 mm}
\def\ro{r_{0}}
\begin{tabular}{||ccc|cc||ccc|cc||cc|cc||ccc|c\TS||} \hline\hline
$\ng$ & $\na$ & $\c$ & $r$&$\ro$ & $\ng$ & $\na$ & $\c$ & $r$&$\ro\!$ &
$\ng$ & $\!p_{{1}}/2\!$& $r$ & $\ro\!$ &
$\ng$ & $p_{{01}}$&$p_{{10}}$ & $r$ \TL\\ \hline
84 & 0 &$-$168 & 1&1 &   0 & 84 &168 & 4&4 &   9 & 3 & 3&3 &   3 & 3 & 1 &4\\
40 & 4 & $-$72 & 2&2 &   4 & 40 & 72 & 3&3 &   21& 1 & 2&2 &   3 & 1 & 3 &4\\
36 & 0 & $-$72 & 2&2 &   0 & 36 & 72 & 4&4 &   9 & 1 & 3&3 &   0 & 3 & 0 &5\\
27 & 3 & $-$48 & 3&4 &   3 & 27 & 48 & 3&4 &   3 & 1 & 3&3 &   0 & 0 & 3 &5\\
24 & 12& $-$24 & 2&2 &   12& 24 & 24 & 3&3 &     &   &  &  &   12& 1 & 0 &3\\
18 & 6 & $-$24 & 3&3 &   6 & 18 & 24 & 3&3 &   13& 0 & 3&3 &   12& 0 & 1 &3\\
16 & 4 & $-$24 & 3&3 &   4 & 16 & 24 & 4&4 &   9 & 0 & 3&4 &   6 & 1 & 0 &4\\
12 & 0 & $-$24 & 3&4 &   0 & 12 & 24 & 4&5 &   7 & 0 & 4&4 &   6 & 0 & 1
&4\TL\\
\hline\hline\end{tabular}\bsk \end{table}

The results of the computation for the model $(1)^9$
are listed in table~2. Although we find more
than $10^5$ inequivalent models, we only end up with 31 different spectra
(all of which already appear in other tensor products). In the first
two columns we list the mirror pairs of the chiral spectra, whereas
columns 3 and 4 contain the symmetric and the asymmetric spectra with
vanishing Euler number, respectively.
We also list the minimal rank for which a given
spectrum  occurs with or without torsion.
Note that, in this model,
only the asymmetric spectra actually require torsion.
Considering the sets $S_r$ of all spectra that occur for groups
of rank $r$ and arbitrary allowed torsions, we find
that $S_r\subset S_{r'}$ if and only if $r\le r'$.
Without torsion, this is true
only up to rank 4, where we have 22 of the 23 symmetric
spectra. At that point we must, according to the well-known construction of
the mirror model for diagonal invariants~\cite{gp},
switch to the mirror spectra, i.e. exchange the numbers $\ng$ and $\na$.
Indeed, when we go from rank 4 to rank 5,
we trade the spectrum $(12,0;-24)$ for its mirror partner, whereas all other
spectra occur in a mirror-symmetric way; beyond rank 5 the set of spectra
decreases exactly the way it should if we use only trivial torsions.

Summing up, we
observe that in both cases, with and without torsion,
all respective spectra can be obtained with groups of rank 1 to 5.
When torsions are involved, these groups
only amount to a small fraction of the inequivalent models.
This peculiar behaviour calls for an explanation. An obvious idea
is that the groups with increasing rank could become redundant at some point.
This cannot be true in the strict sense, of course, because the models
correspond
to different modular invariants (see ref.~\cite{ks}). As the chiral rings,
however, seem to be equivalent -- at least as far as the charge degeneracies
are concerned -- this could be related to a redundancy of the group actions
that only concerns the chiral rings.

Let us call a group $\cg$ {\it reducible} iff there is a strict subgroup $\ch$
such that each state in the (untwisted) chiral ring is invariant under $\cg$
if it is invariant under $\ch$, i.e. iff the untwisted chiral primary fields
of the orbifolds defined by $\cg$ and $\ch$ coincide.%
\footnote{This is, of course, independent of the torsions.}
For example, the `smallest' group $\cg_1^{(3)}$ of rank 3,
\beq  \cg_1^{(3)} = {\pmatrix{ 1\,0\,0\; 1\,1\,1\,1\,1\,0\cr
      0\,1\,0\; 0\,0\,0\,0\,0\,2\cr  0\,0\,1\; 0\,0\,0\,0\,0\,2\cr}},\eeq
which is the only reducible one of this rank,
yields the same projection on the chiral ring as its subgroup of rank 2
\beq \cg_2^{(2)}={\pmatrix{1\,0\,0\;1\,1\,1\,1\,1\,0\cr
  0\,1\,1\;0\,0\,0\,0\,0\,1\cr}}.\eql{g32}
It is indeed easy to check by inspection that the 20 inequivalent groups
with rank larger than 5 are all reducible in this sense.
Furthermore, none of the groups in our list that contribute spectra
that are not present at lower rank is reducible.
This suggests the conjecture that, at least for (2,2) vacua, all reducible
groups can be omitted if we are only interested in computing the set of
(non-singlet) spectra.
In the absence of a proof, such a rule appears plausible only if we find a
mechanism to explain the spectra of a reducible group in terms of its
subgroups.
Having in mind the modding of `quantum symmetries',%
\footnote{These symmetries act only on the twisted sectors, with phases that
   are consistent with the orbifold selection rules~\cite{q}; they
   can be modded using discrete torsions (for examples see ref.~\cite{fk}).}
which can undo part of the original orbifolding, it is conceivable that
a similar mechanism could be at work in the present situation.
This would mean that we should be able to understand the spectrum
of a reducible group only in terms of the  subgroup of rank $r-1$ that
still implements the full projection, and of the subgroups of rank $r-2$
of that group.
The \naive\ guess that these subgroups
(corresponding to a single reduction of the group) should yield all the
spectra has counter-examples at rank 4. It turns out, however, that all
counter-examples that we found allow more than one reduction; the union of
the sets of spectra arising from different reductions indeed gave the
complete set of spectra.

With the methods we developed in this paper it should be straightforward to
completely enumerate more general, and phenomenologically more promising,
classes of string vacua. In particular, these would be general
(2,2) and (2,1) \LGO s, or maybe even all (2,0) ADE models.
Unfortunately, for exceptional invariants, which exist for tensor products
at certain levels~\cite{ex,fksv}, a complete classification is still missing.
Disregarding massless $E_6$ singlets, however, there are no known
spectra that can come only from such exceptional invariants.%
\footnote{On the other hand, there are examples where the full massless
   spectrum is different~\cite{ex}.}

   {\it Acknowledgements.}
It is a pleasure to thank Bert Schellekens for helpful discussions
and comments.

\ve

\pagebreak{\Large\bf Figures} \bsk\bsk

\def\figsca{\funit=0.15truemm \wi=1000 \he=520 \vmul=2 }
\def\figlab{\hlab{-200} \hlab{200} \hlab{-400} \hlab{400}
	    \hlab{-600} \hlab{600} \hlab{-800} \hlab{800}
	    \vlab{100}  \vlab{200} \vlab{300}  \vlab{400} \vmark{500} }
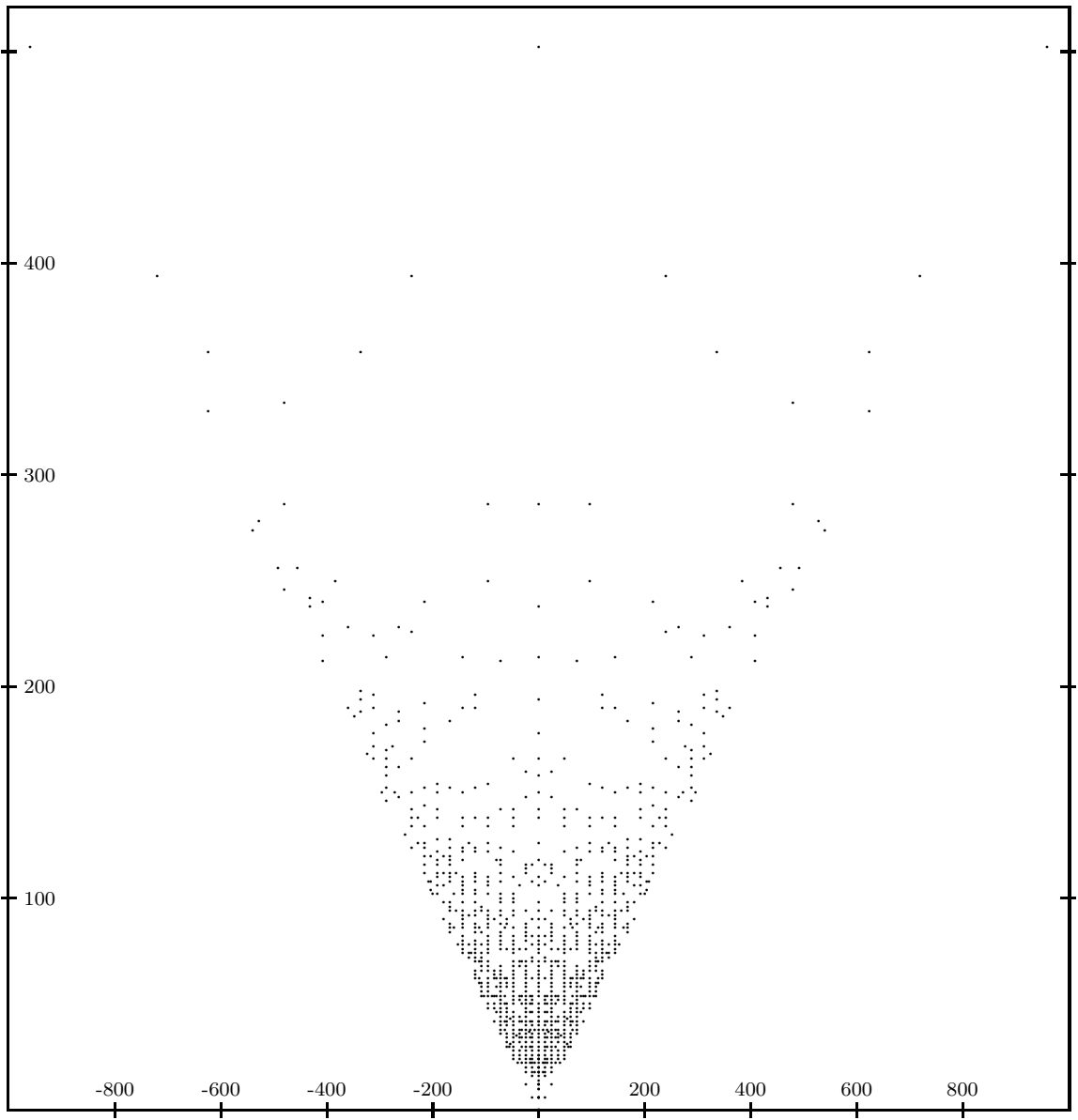
\def\figcap{$\ng+\na$ vs. Euler number $\chi$ for all 943 spectra.}
\Vplo{
\.491,11)\.377,17)\.321,9)\.335,23)\.272,2)\.271,7)\.251,5)\.243,3)\.263,23)
\.287,47)\.242,14)\.227,11)\.229,13)\.208,4)\.214,10)\.222,18)\.221,29)
\.185,5)\.204,24)\.180,6)\.178,10)\.181,13)\.183,15)\.263,95)\.165,3)\.161,5)
\.164,8)\.167,11)\.173,17)\.176,20)\.190,34)\.149,1)\.145,1)\.148,4)\.151,7)
\.153,9)\.155,11)\.157,13)\.163,19)\.179,35)\.155,17)\.143,7)\.140,8)\.147,15)
\.158,26)\.160,28)\.180,48)\.128,2)\.122,2)\.127,7)\.129,9)\.131,11)\.135,15)
\.143,23)\.173,53)\.257,137)\.120,6)\.126,12)\.110,2)\.112,4)\.114,6)\.116,8)
\.117,9)\.121,13)\.126,18)\.130,22)\.141,33)\.144,36)\.150,42)\.174,66)
\.106,2)\.103,1)\.105,3)\.111,9)\.101,1)\.99,3)\.101,5)\.103,7)\.104,8)
\.106,10)\.107,11)\.109,13)\.112,16)\.117,21)\.119,23)\.123,27)\.125,29)
\.90,0)\.94,4)\.98,8)\.101,11)\.105,15)\.84,0)\.85,1)\.86,2)\.89,5)\.90,6)
\.91,7)\.96,12)\.97,13)\.98,14)\.100,16)\.101,17)\.102,18)\.104,20)\.106,22)
\.118,34)\.134,50)\.83,3)\.91,11)\.86,8)\.95,17)\.77,1)\.73,1)\.74,2)\.75,3)
\.76,4)\.77,5)\.79,7)\.80,8)\.81,9)\.82,10)\.83,11)\.85,13)\.87,15)\.88,16)
\.89,17)\.90,18)\.91,19)\.95,23)\.97,25)\.98,26)\.103,31)\.105,33)\.111,39)
\.131,59)\.143,71)\.69,3)\.70,4)\.72,6)\.79,13)\.96,30)\.69,5)\.61,1)\.62,2)
\.63,3)\.65,5)\.66,6)\.67,7)\.68,8)\.69,9)\.70,10)\.72,12)\.73,13)\.74,14)
\.76,16)\.77,17)\.79,19)\.81,21)\.82,22)\.84,24)\.85,25)\.86,26)\.91,31)
\.92,32)\.99,39)\.106,46)\.125,65)\.128,68)\.58,2)\.59,3)\.63,7)\.71,15)
\.54,0)\.55,1)\.56,2)\.57,3)\.60,6)\.61,7)\.62,8)\.63,9)\.64,10)\.66,12)
\.69,15)\.74,20)\.75,21)\.53,1)\.48,0)\.49,1)\.51,3)\.52,4)\.53,5)\.54,6)
\.55,7)\.57,9)\.58,10)\.59,11)\.61,13)\.62,14)\.63,15)\.64,16)\.65,17)\.67,19)
\.68,20)\.69,21)\.70,22)\.71,23)\.75,27)\.77,29)\.78,30)\.79,31)\.85,37)
\.87,39)\.91,43)\.93,45)\.101,53)\.149,101)\.167,119)\.49,5)\.42,0)\.45,3)
\.46,4)\.48,6)\.52,10)\.53,11)\.54,12)\.66,24)\.43,3)\.47,7)\.49,9)\.51,11)
\.79,39)\.36,0)\.37,1)\.38,2)\.39,3)\.40,4)\.41,5)\.42,6)\.43,7)\.44,8)
\.45,9)\.46,10)\.47,11)\.48,12)\.49,13)\.50,14)\.51,15)\.52,16)\.56,20)
\.57,21)\.58,22)\.59,23)\.60,24)\.62,26)\.64,28)\.65,29)\.67,31)\.68,32)
\.69,33)\.72,36)\.74,38)\.76,40)\.77,41)\.80,44)\.89,53)\.124,88)\.35,3)
\.37,5)\.41,9)\.43,11)\.47,15)\.59,27)\.30,0)\.31,1)\.32,2)\.33,3)\.35,5)
\.36,6)\.37,7)\.38,8)\.39,9)\.40,10)\.44,14)\.45,15)\.46,16)\.53,23)\.59,29)
\.60,30)\.29,1)\.29,2)\.35,8)\.24,0)\.25,1)\.26,2)\.27,3)\.28,4)\.29,5)
\.30,6)\.31,7)\.32,8)\.33,9)\.34,10)\.35,11)\.37,13)\.38,14)\.39,15)\.40,16)
\.41,17)\.42,18)\.43,19)\.44,20)\.45,21)\.46,22)\.47,23)\.49,25)\.50,26)
\.51,27)\.52,28)\.53,29)\.55,31)\.58,34)\.59,35)\.61,37)\.62,38)\.63,39)
\.67,43)\.73,49)\.79,55)\.81,57)\.83,59)\.95,71)\.28,7)\.21,1)\.21,3)\.22,4)
\.23,5)\.25,7)\.26,8)\.27,9)\.28,10)\.30,12)\.31,13)\.35,17)\.36,18)\.43,25)
\.44,26)\.47,29)\.62,44)\.19,3)\.23,7)\.25,9)\.27,11)\.29,13)\.33,17)\.35,19)
\.43,27)\.12,0)\.15,3)\.16,4)\.17,5)\.18,6)\.19,7)\.20,8)\.21,9)\.22,10)
\.23,11)\.24,12)\.25,13)\.26,14)\.27,15)\.28,16)\.29,17)\.30,18)\.31,19)
\.32,20)\.33,21)\.34,22)\.35,23)\.36,24)\.37,25)\.38,26)\.39,27)\.40,28)
\.41,29)\.44,32)\.46,34)\.47,35)\.48,36)\.49,37)\.50,38)\.53,41)\.62,50)
\.63,51)\.64,52)\.80,68)\.86,74)\.23,14)\.15,7)\.16,8)\.19,11)\.21,13)\.23,15)
\.25,17)\.27,19)\.29,21)\.31,23)\.6,0)\.11,5)\.12,6)\.13,7)\.14,8)\.15,9)
\.16,10)\.17,11)\.18,12)\.19,13)\.20,14)\.21,15)\.23,17)\.24,18)\.26,20)
\.28,22)\.29,23)\.30,24)\.34,28)\.38,32)\.39,33)\.42,36)\.44,38)\.48,42)
\.57,51)\.61,55)\.11,7)\.13,9)\.21,17)\.3,3)\.5,5)\.6,6)\.7,7)\.8,8)\.9,9)
\.10,10)\.11,11)\.12,12)\.13,13)\.14,14)\.15,15)\.16,16)\.17,17)\.18,18)
\.19,19)\.20,20)\.21,21)\.22,22)\.23,23)\.24,24)\.25,25)\.27,27)\.28,28)
\.29,29)\.30,30)\.31,31)\.32,32)\.33,33)\.34,34)\.35,35)\.37,37)\.38,38)
\.39,39)\.40,40)\.41,41)\.43,43)\.45,45)\.47,47)\.49,49)\.53,53)\.55,55)
\.59,59)\.63,63)\.69,69)\.71,71)\.75,75)\.79,79)\.83,83)\.89,89)\.97,97)
\.107,107)\.119,119)\.143,143)\.251,251)\.7,11)\.9,13)\.17,21)\.0,6)\.5,11)
\.6,12)\.7,13)\.8,14)\.9,15)\.10,16)\.11,17)\.12,18)\.13,19)\.14,20)\.15,21)
\.17,23)\.18,24)\.20,26)\.22,28)\.23,29)\.24,30)\.28,34)\.32,38)\.33,39)
\.36,42)\.38,44)\.42,48)\.51,57)\.55,61)\.7,15)\.8,16)\.11,19)\.13,21)\.15,23)
\.17,25)\.19,27)\.21,29)\.23,31)\.14,23)\.0,12)\.3,15)\.4,16)\.5,17)\.6,18)
\.7,19)\.8,20)\.9,21)\.10,22)\.11,23)\.12,24)\.13,25)\.14,26)\.15,27)\.16,28)
\.17,29)\.18,30)\.19,31)\.20,32)\.21,33)\.22,34)\.23,35)\.24,36)\.25,37)
\.26,38)\.27,39)\.28,40)\.29,41)\.32,44)\.34,46)\.35,47)\.36,48)\.37,49)
\.38,50)\.41,53)\.50,62)\.51,63)\.52,64)\.68,80)\.74,86)\.3,19)\.7,23)\.9,25)
\.11,27)\.13,29)\.17,33)\.19,35)\.27,43)\.3,21)\.4,22)\.5,23)\.7,25)\.8,26)
\.9,27)\.10,28)\.12,30)\.13,31)\.17,35)\.18,36)\.25,43)\.26,44)\.29,47)
\.44,62)\.1,21)\.7,28)\.0,24)\.1,25)\.2,26)\.3,27)\.4,28)\.5,29)\.6,30)
\.7,31)\.8,32)\.9,33)\.10,34)\.11,35)\.13,37)\.14,38)\.15,39)\.16,40)\.17,41)
\.18,42)\.19,43)\.20,44)\.21,45)\.22,46)\.23,47)\.25,49)\.26,50)\.27,51)
\.28,52)\.29,53)\.31,55)\.34,58)\.35,59)\.37,61)\.38,62)\.39,63)\.43,67)
\.49,73)\.55,79)\.57,81)\.59,83)\.71,95)\.2,29)\.8,35)\.1,29)\.0,30)\.1,31)
\.2,32)\.3,33)\.5,35)\.6,36)\.7,37)\.8,38)\.9,39)\.10,40)\.14,44)\.15,45)
\.16,46)\.23,53)\.29,59)\.30,60)\.3,35)\.5,37)\.9,41)\.11,43)\.15,47)\.27,59)
\.0,36)\.1,37)\.2,38)\.3,39)\.4,40)\.5,41)\.6,42)\.7,43)\.8,44)\.9,45)\.10,46)
\.11,47)\.12,48)\.13,49)\.14,50)\.15,51)\.16,52)\.20,56)\.21,57)\.22,58)
\.23,59)\.24,60)\.26,62)\.28,64)\.29,65)\.31,67)\.32,68)\.33,69)\.36,72)
\.38,74)\.40,76)\.41,77)\.44,80)\.53,89)\.88,124)\.3,43)\.7,47)\.9,49)\.11,51)
\.39,79)\.0,42)\.3,45)\.4,46)\.6,48)\.10,52)\.11,53)\.12,54)\.24,66)\.5,49)
\.0,48)\.1,49)\.3,51)\.4,52)\.5,53)\.6,54)\.7,55)\.9,57)\.10,58)\.11,59)
\.13,61)\.14,62)\.15,63)\.16,64)\.17,65)\.19,67)\.20,68)\.21,69)\.22,70)
\.23,71)\.27,75)\.29,77)\.30,78)\.31,79)\.37,85)\.39,87)\.43,91)\.45,93)
\.53,101)\.101,149)\.119,167)\.1,53)\.0,54)\.1,55)\.2,56)\.3,57)\.6,60)
\.7,61)\.8,62)\.9,63)\.10,64)\.12,66)\.15,69)\.20,74)\.21,75)\.2,58)\.3,59)
\.7,63)\.15,71)\.1,61)\.2,62)\.3,63)\.5,65)\.6,66)\.7,67)\.8,68)\.9,69)
\.10,70)\.12,72)\.13,73)\.14,74)\.16,76)\.17,77)\.19,79)\.21,81)\.22,82)
\.24,84)\.25,85)\.26,86)\.31,91)\.32,92)\.39,99)\.46,106)\.65,125)\.68,128)
\.5,69)\.3,69)\.4,70)\.6,72)\.13,79)\.30,96)\.1,73)\.2,74)\.3,75)\.4,76)
\.5,77)\.7,79)\.8,80)\.9,81)\.10,82)\.11,83)\.13,85)\.15,87)\.16,88)\.17,89)
\.18,90)\.19,91)\.23,95)\.25,97)\.26,98)\.31,103)\.33,105)\.39,111)\.59,131)
\.71,143)\.1,77)\.8,86)\.17,95)\.3,83)\.11,91)\.0,84)\.1,85)\.2,86)\.5,89)
\.6,90)\.7,91)\.12,96)\.13,97)\.14,98)\.16,100)\.17,101)\.18,102)\.20,104)
\.22,106)\.34,118)\.50,134)\.0,90)\.4,94)\.8,98)\.11,101)\.15,105)\.3,99)
\.5,101)\.7,103)\.8,104)\.10,106)\.11,107)\.13,109)\.16,112)\.21,117)\.23,119)
\.27,123)\.29,125)\.1,101)\.1,103)\.3,105)\.9,111)\.2,106)\.2,110)\.4,112)
\.6,114)\.8,116)\.9,117)\.13,121)\.18,126)\.22,130)\.33,141)\.36,144)\.42,150)
\.66,174)\.6,120)\.12,126)\.2,122)\.7,127)\.9,129)\.11,131)\.15,135)\.23,143)
\.53,173)\.137,257)\.2,128)\.8,140)\.15,147)\.26,158)\.28,160)\.48,180)
\.7,143)\.17,155)\.1,145)\.4,148)\.7,151)\.9,153)\.11,155)\.13,157)\.19,163)
\.35,179)\.1,149)\.5,161)\.8,164)\.11,167)\.17,173)\.20,176)\.34,190)\.3,165)
\.10,178)\.13,181)\.15,183)\.95,263)\.6,180)\.5,185)\.24,204)\.29,221)\.4,208)
\.10,214)\.18,222)\.11,227)\.13,229)\.14,242)\.3,243)\.23,263)\.47,287)
\.5,251)\.7,271)\.2,272)\.9,321)\.23,335)\.17,377)\.11,491)\.6,6)\.10,10)
\.12,12)\.0,0)\.6,6)\.10,10)\.12,12)\.3,3)\.5,5)\.9,9)\.13,13)\.21,21)\.3,3)
\.0,0)\.3,3)\.9,9) 
}

\def\figsca{\funit=.7truemm \wi=240 \he=120 \vmul=1}
\def\figlab{\hlab{-60} \hlab{60} \hlab{-120} \hlab{120} \hlab{-180} \hlab{180}
            \vlab{30}  \vlab{60} \vlab{90} }
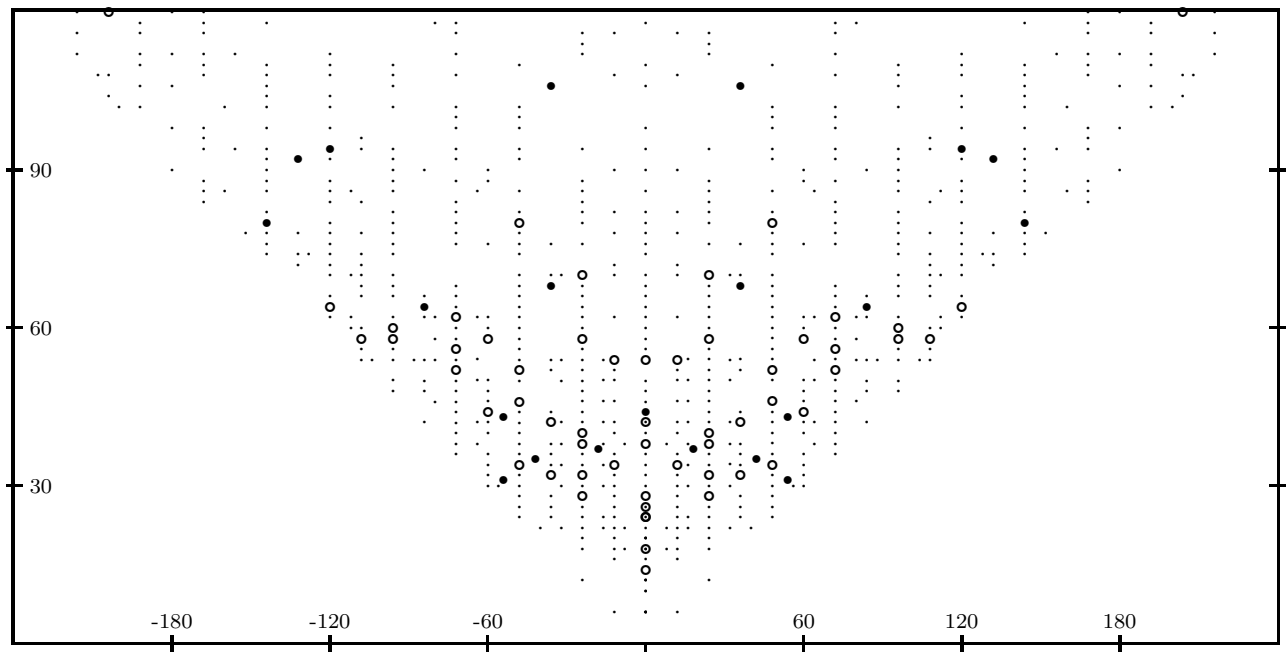
\def\figcap{$\ng+\na$ vs. Euler number for
  the spectra that can come [only] from $E_7$ invariants (large dots
[circles]);
  the remaining spectra with $\ng+\na\le120$ are indicated by small dots. }
\Vplo{
\def\npt{{\funit=1pt \circle*1}} 
\.110,2)\.112,4)\.114,6)\.106,2)\.103,1)\.105,3)\.101,1)\.99,3)\.101,5)
\.103,7)\.104,8)\.106,10)\.107,11)\.90,0)\.94,4)\.98,8)\.101,11)\.105,15)
\.84,0)\.85,1)\.86,2)\.89,5)\.90,6)\.91,7)\.96,12)\.97,13)\.98,14)\.100,16)
\.101,17)\.102,18)\.83,3)\.91,11)\.86,8)\.95,17)\.77,1)\.73,1)\.74,2)\.75,3)
\.77,5)\.79,7)\.80,8)\.81,9)\.82,10)\.83,11)\.85,13)\.87,15)\.88,16)\.89,17)
\.90,18)\.91,19)\.95,23)\.69,3)\.70,4)\.72,6)\.69,5)\.61,1)\.63,3)\.65,5)
\.66,6)\.67,7)\.68,8)\.69,9)\.70,10)\.72,12)\.73,13)\.74,14)\.76,16)\.79,19)
\.81,21)\.82,22)\.84,24)\.85,25)\.86,26)\.58,2)\.59,3)\.63,7)\.71,15)\.54,0)
\.55,1)\.57,3)\.60,6)\.61,7)\.62,8)\.63,9)\.64,10)\.66,12)\.69,15)\.74,20)
\.75,21)\.53,1)\.48,0)\.49,1)\.51,3)\.52,4)\.55,7)\.57,9)\.58,10)\.59,11)
\.61,13)\.62,14)\.63,15)\.64,16)\.65,17)\.67,19)\.68,20)\.69,21)\.70,22)
\.71,23)\.75,27)\.77,29)\.78,30)\.79,31)\.49,5)\.42,0)\.45,3)\.46,4)\.48,6)
\.52,10)\.54,12)\.66,24)\.43,3)\.47,7)\.49,9)\.51,11)\.79,39)\.36,0)\.37,1)
\.38,2)\.39,3)\.40,4)\.41,5)\.42,6)\.43,7)\.45,9)\.47,11)\.48,12)\.50,14)
\.51,15)\.52,16)\.56,20)\.57,21)\.58,22)\.59,23)\.60,24)\.62,26)\.64,28)
\.65,29)\.67,31)\.68,32)\.69,33)\.72,36)\.74,38)\.76,40)\.77,41)\.35,3)
\.37,5)\.41,9)\.43,11)\.47,15)\.59,27)\.30,0)\.31,1)\.32,2)\.33,3)\.35,5)
\.36,6)\.38,8)\.39,9)\.40,10)\.45,15)\.46,16)\.53,23)\.59,29)\.60,30)\.29,1)
\.24,0)\.25,1)\.26,2)\.27,3)\.28,4)\.30,6)\.31,7)\.32,8)\.33,9)\.34,10)
\.37,13)\.39,15)\.40,16)\.41,17)\.42,18)\.43,19)\.44,20)\.45,21)\.46,22)
\.47,23)\.49,25)\.50,26)\.51,27)\.53,29)\.55,31)\.58,34)\.59,35)\.61,37)
\.62,38)\.63,39)\.67,43)\.21,1)\.21,3)\.22,4)\.23,5)\.26,8)\.27,9)\.28,10)
\.31,13)\.35,17)\.36,18)\.44,26)\.47,29)\.19,3)\.23,7)\.25,9)\.27,11)\.29,13)
\.33,17)\.35,19)\.43,27)\.12,0)\.15,3)\.16,4)\.17,5)\.18,6)\.19,7)\.21,9)
\.23,11)\.24,12)\.27,15)\.28,16)\.29,17)\.30,18)\.31,19)\.32,20)\.33,21)
\.34,22)\.36,24)\.37,25)\.38,26)\.39,27)\.40,28)\.44,32)\.46,34)\.47,35)
\.48,36)\.49,37)\.50,38)\.53,41)\.62,50)\.63,51)\.64,52)\.15,7)\.16,8)\.19,11)
\.21,13)\.23,15)\.25,17)\.27,19)\.29,21)\.31,23)\.6,0)\.11,5)\.12,6)\.13,7)
\.14,8)\.15,9)\.16,10)\.17,11)\.18,12)\.19,13)\.21,15)\.23,17)\.24,18)\.26,20)
\.28,22)\.29,23)\.34,28)\.38,32)\.39,33)\.42,36)\.44,38)\.48,42)\.57,51)
\.61,55)\.11,7)\.13,9)\.21,17)\.3,3)\.5,5)\.6,6)\.8,8)\.9,9)\.10,10)\.11,11)
\.12,12)\.15,15)\.16,16)\.17,17)\.18,18)\.20,20)\.21,21)\.23,23)\.24,24)
\.25,25)\.28,28)\.29,29)\.30,30)\.31,31)\.32,32)\.33,33)\.34,34)\.35,35)
\.37,37)\.38,38)\.39,39)\.40,40)\.41,41)\.43,43)\.45,45)\.47,47)\.49,49)
\.53,53)\.55,55)\.59,59)\.7,11)\.9,13)\.17,21)\.0,6)\.5,11)\.6,12)\.7,13)
\.8,14)\.9,15)\.10,16)\.11,17)\.12,18)\.13,19)\.15,21)\.17,23)\.18,24)\.20,26)
\.22,28)\.23,29)\.28,34)\.32,38)\.33,39)\.36,42)\.38,44)\.42,48)\.51,57)
\.55,61)\.7,15)\.8,16)\.11,19)\.13,21)\.15,23)\.17,25)\.19,27)\.21,29)\.23,31)
\.0,12)\.3,15)\.4,16)\.5,17)\.6,18)\.7,19)\.9,21)\.11,23)\.12,24)\.15,27)
\.16,28)\.17,29)\.18,30)\.19,31)\.20,32)\.21,33)\.22,34)\.24,36)\.25,37)
\.26,38)\.27,39)\.28,40)\.32,44)\.34,46)\.35,47)\.36,48)\.37,49)\.38,50)
\.41,53)\.50,62)\.51,63)\.52,64)\.3,19)\.7,23)\.9,25)\.11,27)\.13,29)\.17,33)
\.19,35)\.27,43)\.3,21)\.4,22)\.5,23)\.8,26)\.9,27)\.10,28)\.13,31)\.17,35)
\.18,36)\.26,44)\.29,47)\.1,21)\.0,24)\.1,25)\.2,26)\.3,27)\.4,28)\.6,30)
\.7,31)\.8,32)\.9,33)\.10,34)\.13,37)\.15,39)\.16,40)\.17,41)\.18,42)\.19,43)
\.20,44)\.21,45)\.22,46)\.23,47)\.25,49)\.26,50)\.27,51)\.29,53)\.31,55)
\.34,58)\.35,59)\.37,61)\.38,62)\.39,63)\.43,67)\.1,29)\.0,30)\.1,31)\.2,32)
\.3,33)\.5,35)\.6,36)\.8,38)\.9,39)\.10,40)\.15,45)\.16,46)\.23,53)\.29,59)
\.30,60)\.3,35)\.5,37)\.9,41)\.11,43)\.15,47)\.27,59)\.0,36)\.1,37)\.2,38)
\.3,39)\.4,40)\.5,41)\.6,42)\.7,43)\.9,45)\.11,47)\.12,48)\.14,50)\.15,51)
\.16,52)\.20,56)\.21,57)\.22,58)\.23,59)\.24,60)\.26,62)\.28,64)\.29,65)
\.31,67)\.32,68)\.33,69)\.36,72)\.38,74)\.40,76)\.41,77)\.3,43)\.7,47)\.9,49)
\.11,51)\.39,79)\.0,42)\.3,45)\.4,46)\.6,48)\.10,52)\.12,54)\.24,66)\.5,49)
\.0,48)\.1,49)\.3,51)\.4,52)\.7,55)\.9,57)\.10,58)\.11,59)\.13,61)\.14,62)
\.15,63)\.16,64)\.17,65)\.19,67)\.20,68)\.21,69)\.22,70)\.23,71)\.27,75)
\.29,77)\.30,78)\.31,79)\.1,53)\.0,54)\.1,55)\.3,57)\.6,60)\.7,61)\.8,62)
\.9,63)\.10,64)\.12,66)\.15,69)\.20,74)\.21,75)\.2,58)\.3,59)\.7,63)\.15,71)
\.1,61)\.3,63)\.5,65)\.6,66)\.7,67)\.8,68)\.9,69)\.10,70)\.12,72)\.13,73)
\.14,74)\.16,76)\.19,79)\.21,81)\.22,82)\.24,84)\.25,85)\.26,86)\.5,69)
\.3,69)\.4,70)\.6,72)\.1,73)\.2,74)\.3,75)\.5,77)\.7,79)\.8,80)\.9,81)\.10,82)
\.11,83)\.13,85)\.15,87)\.16,88)\.17,89)\.18,90)\.19,91)\.23,95)\.1,77)
\.8,86)\.17,95)\.3,83)\.11,91)\.0,84)\.1,85)\.2,86)\.5,89)\.6,90)\.7,91)
\.12,96)\.13,97)\.14,98)\.16,100)\.17,101)\.18,102)\.0,90)\.4,94)\.8,98)
\.11,101)\.15,105)\.3,99)\.5,101)\.7,103)\.8,104)\.10,106)\.11,107)\.1,101)
\.1,103)\.3,105)\.2,106)\.2,110)\.4,112)\.6,114)\.6,6)\.10,10)\.0,0)\.6,6)
\.10,10)\.3,3)\.5,5)\.13,13)\.3,3)\.0,0)\.3,3)\.9,9)
%
\def\npt{\funit=1pt\circle3}  
\.111,9)\.62,2)\.56,2)\.53,5)\.54,6)\.44,8)\.46,10)\.49,13)\.37,7)\.44,14)
\.29,5)\.35,11)\.38,14)\.52,28)\.25,7)\.30,12)\.20,8)\.22,10)\.25,13)\.26,14)
\.35,23)\.41,29)\.20,14)\.30,24)\.7,7)\.13,13)\.14,14)\.19,19)\.27,27)\.14,20)
\.24,30)\.8,20)\.10,22)\.13,25)\.14,26)\.23,35)\.29,41)\.7,25)\.12,30)\.5,29)
\.11,35)\.14,38)\.28,52)\.7,37)\.14,44)\.8,44)\.10,46)\.13,49)\.5,53)\.6,54)
\.2,56)\.2,62)\.9,111)\.12,12)\.12,12)\.9,9)\.21,21) 
\def\npt{\funit=1pt\circle*3} 
\.76,4)\.79,13)\.77,17)\.53,11)\.29,2)\.35,8)\.28,7)\.43,25)\.62,44)\.23,14)
\.22,22)\.14,23)\.25,43)\.44,62)\.7,28)\.2,29)\.8,35)\.11,53)\.17,77)\.13,79)
\.4,76) 
}

\vfill

\def\figsca{\funit=.7truemm \wi=240 \he=120 \vmul=1}
\def\figlab{\hlab{-60} \hlab{60} \hlab{-120} \hlab{120} \hlab{-180} \hlab{180}
            \vlab{30}  \vlab{60} \vlab{90} }
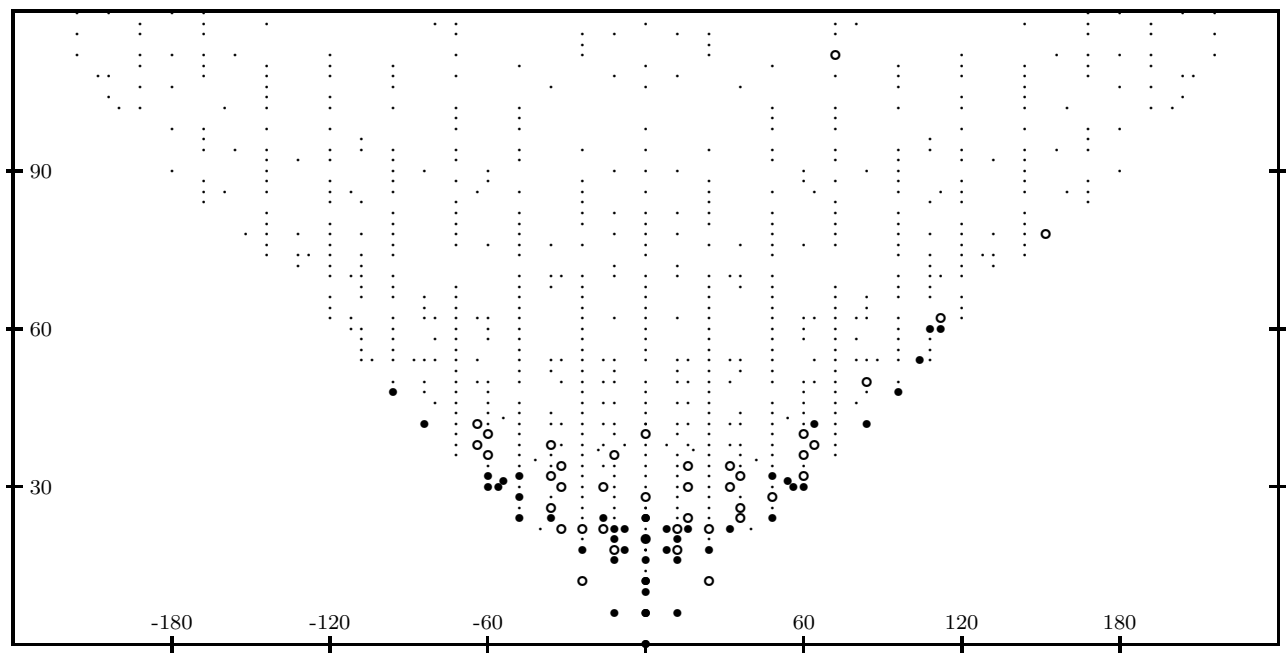
\def\figcap{$\ng+\na$ vs. Euler number for spectra not occurring for any
Landau--Ginzburg/ADE \\
orbifold without torsion (large dots/circles), and all others with
$\ng+\na\le120$ (small dots).}
\Vplo{
\def\npt{{\funit=1pt \circle*1}} 
\.110,2)\.112,4)\.114,6)\.106,2)\.103,1)\.105,3)\.111,9)\.101,1)\.99,3)
\.101,5)\.103,7)\.104,8)\.106,10)\.107,11)\.90,0)\.94,4)\.98,8)\.101,11)
\.105,15)\.84,0)\.85,1)\.86,2)\.89,5)\.90,6)\.91,7)\.96,12)\.97,13)\.98,14)
\.100,16)\.101,17)\.102,18)\.83,3)\.91,11)\.86,8)\.95,17)\.77,1)\.73,1)
\.74,2)\.75,3)\.76,4)\.77,5)\.79,7)\.80,8)\.81,9)\.82,10)\.83,11)\.85,13)
\.87,15)\.88,16)\.89,17)\.90,18)\.91,19)\.95,23)\.69,3)\.70,4)\.72,6)\.79,13)
\.69,5)\.61,1)\.62,2)\.63,3)\.65,5)\.66,6)\.67,7)\.68,8)\.69,9)\.70,10)
\.72,12)\.73,13)\.74,14)\.76,16)\.77,17)\.79,19)\.81,21)\.82,22)\.84,24)
\.85,25)\.86,26)\.58,2)\.59,3)\.63,7)\.71,15)\.54,0)\.55,1)\.56,2)\.57,3)
\.60,6)\.61,7)\.62,8)\.63,9)\.64,10)\.66,12)\.69,15)\.74,20)\.75,21)\.53,1)
\.49,1)\.51,3)\.52,4)\.53,5)\.54,6)\.55,7)\.57,9)\.58,10)\.59,11)\.61,13)
\.62,14)\.63,15)\.64,16)\.65,17)\.67,19)\.68,20)\.69,21)\.70,22)\.71,23)
\.75,27)\.77,29)\.78,30)\.79,31)\.49,5)\.45,3)\.46,4)\.48,6)\.52,10)\.53,11)
\.54,12)\.66,24)\.43,3)\.47,7)\.49,9)\.51,11)\.79,39)\.36,0)\.37,1)\.38,2)
\.39,3)\.40,4)\.41,5)\.42,6)\.43,7)\.44,8)\.45,9)\.46,10)\.47,11)\.48,12)
\.49,13)\.50,14)\.51,15)\.52,16)\.56,20)\.57,21)\.58,22)\.59,23)\.60,24)
\.62,26)\.64,28)\.65,29)\.67,31)\.68,32)\.69,33)\.72,36)\.74,38)\.76,40)
\.77,41)\.41,9)\.43,11)\.47,15)\.59,27)\.32,2)\.36,6)\.37,7)\.38,8)\.39,9)
\.40,10)\.44,14)\.45,15)\.46,16)\.53,23)\.59,29)\.60,30)\.35,8)\.25,1)\.27,3)
\.29,5)\.30,6)\.31,7)\.32,8)\.33,9)\.34,10)\.35,11)\.37,13)\.38,14)\.39,15)
\.40,16)\.41,17)\.42,18)\.43,19)\.44,20)\.45,21)\.46,22)\.47,23)\.49,25)
\.50,26)\.51,27)\.52,28)\.53,29)\.55,31)\.58,34)\.59,35)\.61,37)\.62,38)
\.63,39)\.67,43)\.28,7)\.21,1)\.23,5)\.26,8)\.27,9)\.30,12)\.31,13)\.35,17)
\.36,18)\.43,25)\.44,26)\.47,29)\.62,44)\.27,11)\.29,13)\.33,17)\.35,19)
\.43,27)\.16,4)\.18,6)\.19,7)\.20,8)\.21,9)\.22,10)\.23,11)\.24,12)\.25,13)
\.26,14)\.27,15)\.28,16)\.29,17)\.30,18)\.31,19)\.32,20)\.33,21)\.34,22)
\.35,23)\.36,24)\.37,25)\.38,26)\.39,27)\.40,28)\.41,29)\.44,32)\.46,34)
\.47,35)\.48,36)\.49,37)\.50,38)\.53,41)\.62,50)\.63,51)\.64,52)\.23,14)
\.21,13)\.23,15)\.25,17)\.27,19)\.29,21)\.31,23)\.15,9)\.16,10)\.17,11)
\.18,12)\.19,13)\.20,14)\.23,17)\.24,18)\.26,20)\.28,22)\.29,23)\.30,24)
\.34,28)\.38,32)\.39,33)\.42,36)\.44,38)\.48,42)\.57,51)\.61,55)\.21,17)
\.7,7)\.9,9)\.11,11)\.13,13)\.15,15)\.16,16)\.17,17)\.18,18)\.19,19)\.21,21)
\.22,22)\.23,23)\.24,24)\.25,25)\.27,27)\.28,28)\.29,29)\.30,30)\.31,31)
\.32,32)\.33,33)\.34,34)\.35,35)\.37,37)\.38,38)\.39,39)\.40,40)\.41,41)
\.43,43)\.45,45)\.47,47)\.49,49)\.53,53)\.55,55)\.59,59)\.17,21)\.9,15)
\.10,16)\.11,17)\.12,18)\.13,19)\.14,20)\.15,21)\.17,23)\.18,24)\.20,26)
\.22,28)\.23,29)\.24,30)\.28,34)\.32,38)\.33,39)\.36,42)\.38,44)\.42,48)
\.51,57)\.55,61)\.15,23)\.17,25)\.19,27)\.21,29)\.23,31)\.14,23)\.4,16)
\.6,18)\.7,19)\.8,20)\.9,21)\.10,22)\.11,23)\.12,24)\.13,25)\.14,26)\.15,27)
\.16,28)\.17,29)\.18,30)\.19,31)\.20,32)\.21,33)\.22,34)\.23,35)\.24,36)
\.25,37)\.26,38)\.27,39)\.28,40)\.29,41)\.32,44)\.34,46)\.35,47)\.36,48)
\.37,49)\.38,50)\.41,53)\.50,62)\.51,63)\.52,64)\.11,27)\.13,29)\.17,33)
\.19,35)\.27,43)\.5,23)\.8,26)\.9,27)\.10,28)\.12,30)\.13,31)\.17,35)\.18,36)
\.25,43)\.26,44)\.29,47)\.44,62)\.1,21)\.7,28)\.1,25)\.3,27)\.5,29)\.6,30)
\.7,31)\.8,32)\.9,33)\.10,34)\.11,35)\.13,37)\.14,38)\.15,39)\.16,40)\.17,41)
\.18,42)\.19,43)\.20,44)\.21,45)\.22,46)\.23,47)\.25,49)\.26,50)\.27,51)
\.28,52)\.29,53)\.31,55)\.34,58)\.35,59)\.37,61)\.38,62)\.39,63)\.43,67)
\.8,35)\.2,32)\.6,36)\.7,37)\.8,38)\.9,39)\.10,40)\.14,44)\.15,45)\.16,46)
\.23,53)\.29,59)\.30,60)\.9,41)\.11,43)\.15,47)\.27,59)\.0,36)\.1,37)\.2,38)
\.3,39)\.4,40)\.5,41)\.6,42)\.7,43)\.8,44)\.9,45)\.10,46)\.11,47)\.12,48)
\.13,49)\.14,50)\.15,51)\.16,52)\.20,56)\.21,57)\.22,58)\.23,59)\.24,60)
\.26,62)\.28,64)\.29,65)\.31,67)\.32,68)\.33,69)\.36,72)\.40,76)\.41,77)
\.3,43)\.7,47)\.9,49)\.11,51)\.39,79)\.3,45)\.6,48)\.10,52)\.11,53)\.12,54)
\.24,66)\.5,49)\.1,49)\.3,51)\.4,52)\.5,53)\.6,54)\.7,55)\.9,57)\.10,58)
\.11,59)\.13,61)\.14,62)\.15,63)\.16,64)\.17,65)\.19,67)\.20,68)\.21,69)
\.22,70)\.23,71)\.27,75)\.29,77)\.30,78)\.31,79)\.0,54)\.1,55)\.2,56)\.6,60)
\.7,61)\.8,62)\.9,63)\.10,64)\.12,66)\.15,69)\.20,74)\.21,75)\.7,63)\.15,71)
\.1,61)\.2,62)\.3,63)\.5,65)\.6,66)\.7,67)\.8,68)\.9,69)\.10,70)\.12,72)
\.13,73)\.14,74)\.16,76)\.17,77)\.19,79)\.21,81)\.22,82)\.24,84)\.25,85)
\.26,86)\.5,69)\.3,69)\.4,70)\.6,72)\.13,79)\.1,73)\.2,74)\.3,75)\.4,76)
\.5,77)\.7,79)\.8,80)\.9,81)\.10,82)\.11,83)\.13,85)\.15,87)\.16,88)\.17,89)
\.18,90)\.19,91)\.23,95)\.8,86)\.17,95)\.3,83)\.11,91)\.0,84)\.1,85)\.2,86)
\.5,89)\.6,90)\.7,91)\.12,96)\.13,97)\.14,98)\.16,100)\.17,101)\.18,102)
\.0,90)\.4,94)\.8,98)\.11,101)\.15,105)\.3,99)\.5,101)\.7,103)\.8,104)\.10,106)
\.11,107)\.1,101)\.1,103)\.3,105)\.9,111)\.2,106)\.2,110)\.4,112)\.6,114)
\.3,3)\.5,5)\.9,9)\.13,13)\.21,21)\.9,9)
%
\def\npt{\funit=1pt\circle3}  
\.35,3)\.37,5)\.33,3)\.35,5)\.22,4)\.25,7)\.28,10)\.19,3)\.23,7)\.25,9)
\.12,0)\.17,5)\.15,7)\.19,11)\.12,6)\.21,15)\.10,10)\.14,14)\.20,20)\.6,12)
\.8,14)\.8,16)\.11,19)\.13,21)\.0,12)\.5,17)\.7,23)\.9,25)\.3,21)\.4,22)
\.7,25)\.2,26)\.1,31)\.3,33)\.5,35)\.3,35)\.38,74)\.4,46)\.3,59)\.1,77)
%
\def\npt{\funit=1pt\circle*3} 
\.48,0)\.42,0)\.30,0)\.31,1)\.29,1)\.29,2)\.24,0)\.26,2)\.28,4)\.21,3)\.15,3)
\.16,8)\.6,0)\.11,5)\.13,7)\.14,8)\.11,7)\.13,9)\.3,3)\.5,5)\.6,6)\.8,8)
\.12,12)\.7,11)\.9,13)\.0,6)\.5,11)\.7,13)\.7,15)\.3,15)\.3,19)\.0,24)\.4,28)
\.2,29)\.1,29)\.0,30)\.5,37)\.0,42)\.0,48)\.1,53)\.3,57)\.2,58)\.6,6)\.10,10)
\.12,12)\.0,0)\.6,6)\.10,10)\.12,12)\.3,3)\.0,0)\.3,3)
}
\end{document}